\documentstyle[12pt,psfig]{article}
\def\reference{\parskip 0pt\par\noindent\hangindent 0.5 truecm}

\begin{document}
%
%
\title{PSRCHIVE and PSRFITS\\ An Open Approach to Radio Pulsar \\ Data
Storage and Analysis}
%


\author{A. W. Hotan, $^{1,2}$
        W. van Straten, $^{3}$
        R. N. Manchester $^{2}$
} 

\date{}
\maketitle

{\center
$^1$ Swinburne Centre for Astrophysics \& Supercomputing, 
     \\Mail \# 31, PO Box 218, Hawthorn, VIC 3122, Australia
     \\ahotan@astro.swin.edu.au\\[3mm]
$^2$ Australia Telescope National Facility,
     \\PO Box 76, Epping, NSW 1710, Australia, 
     \\Dick.Manchester@csiro.au\\[3mm]
$^3$ Netherlands Foundation for Research in Astronomy,
     \\ PO Box 2, 7990 AA Dwingeloo, The Netherlands,
     \\straten@astron.nl\\[3mm]
}

%
\begin{abstract}

A new set of software applications and libraries for use in the
archival and analysis of pulsar astronomical data is introduced.
Known collectively as the PSRCHIVE scheme, the code was developed in
parallel with a new data storage format called PSRFITS, which is based
on the Flexible Image Transport System (FITS). Both of these projects
utilise a modular, object-oriented design philosophy.  PSRCHIVE is an
open source development environment that incorporates an extensive
range of C++ object classes and pre-built command line and graphical
utilities.  These deal transparently and simultaneously with multiple
data storage formats, thereby enhancing data portability and
facilitating the adoption of the PSRFITS file format.  Here, data are
stored in a series of modular header-data units that provide
flexibility and scope for future expansion.  As it is based on FITS,
various standard libraries and applications may be used for data
input/output and visualisation.  Both PSRCHIVE and PSRFITS are made
publicly available to the academic community in the hope that this
will promote their widespread use and acceptance.

\end{abstract}


\bigskip

%
%

\section{Introduction} 
\subsection{Collaborative scientific software development}

Modern, highly specialised experimental systems often require
extensive original software development. This is true for all tasks
from direct hardware control through to data reduction. Whilst
individual research groups often approach such software development
from an isolated perspective, the proliferation of digital hardware
and wide-area networking makes global cooperative software development
far more attractive, provided suitable common ground exists. Both
cooperative software development and the adoption of standard packages
provide a number of distinct advantages to the research community.
For instance, as less effort is wasted unnecessarily duplicating the
work of others, cooperative development can lead to more efficient
allocation of resources. In addition, supporting the requirements of a
larger user community promotes the development of basic, general
purpose routines that may be used in a wider variety of situations.
These influences result in more modular and extensible software.\\

However, it should be noted that a greater level of care and
cooperation is required in collaborative software development,
especially the open source approach advocated in this paper.  For
example, in contrast to most commercial software, ``black box''
modularity is undesirable in scientific analysis, especially when the
application of certain algorithms requires experienced judgment.
Open, well-documented code provides researchers with an accurate
understanding of third-party analytical tools.  Therefore,
contributing developers must be willing to put their code in the
public domain, making it freely available for non-commercial use by
any other academic organisation.  Although this facilitates the
exchange of ideas, it also raises the issue of potential loss of
intellectual property, which might discourage some authors.\\

It is also the case that collaborative development tends to become
de-centralised, especially when multiple developers have the ability
to commit fundamental changes to the code.  Effective communication
between the core developers becomes essential to the smooth running of
the project, necessitating greater attention to version control, the
maintenance of stable releases, and the development of extensive and
concise documentation.  Also, when a wider user community is affected
by modifications to the software, exhaustive methods must be employed
to ensure the validity of changes and the integrity of the system as a
whole.  Although each of these issues tend to increase the workload of
the collaborative developer, a much larger body of users will benefit
from the effort.

\subsection{Software development in the pulsar community}

The global pulsar community is ideally suited to adopt a collaborative
approach to software development.  It consists of a relatively small
number of locally centralised groups that deal with different
telescopes and instruments, leading to several parallel but
incompatible software development paths.  As each path tends to be
built around a highly specific data storage format, cross examination
of data and algorithms is problematic.  In addition, because such
software is generally designed for a limited purpose, it is often
difficult to extend its functionality without introducing obfuscated
code.  This is especially true when the program develops in an
experimental fashion, as is often the case with scientific
applications.  In order to avoid future inflexibility, sufficient time
and care must be invested during the planning stage, calling on input
from both experienced software developers and pulsar astronomers.

\subsection{Radio pulsar data}

Radio pulsars are broadband point sources of highly polarised emission
that exhibit rapid pulsations with a characteristic period anywhere
between one millisecond and ten seconds.  They are thought to be
rotating neutron stars with a strong dipolar magnetic field whose axis
is not aligned with the rotation axis of the star (Gold 1968).
Intense beams of emission originate at the magnetic poles, which sweep
across the sky with each rotation of the star and produce the pulsed
radio signal observed at Earth.\\

The characteristic \textit{signature} of any radio pulsar is its
integrated polarimetric pulse profile, given by the observed Stokes
parameters averaged (folded) as a function of pulse longitude over
several thousand individual pulses (Helfand, Manchester, \& Taylor
1975). Under the influence of electrons in the interstellar medium
(Taylor \& Cordes 1993), this pulsed signal is broadened by dispersive
frequency smearing, which must be corrected in order to infer the
shape of the characteristic profile at the source.  This is normally
done by dividing the observed bandwidth into narrow frequency
channels, which are appropriately delayed relative to each other
before summing the detected flux densities in each channel.  However,
as the dispersion measure may vary with time or may not be known with
sufficient accuracy at the time of the observation, it is often
necessary to store the individual pulse profiles observed in each
frequency channel.\\

Additionally, it is possible to create a mean pulse profile only if a
suitably accurate model of the pulsar's spin frequency and phase is
available.  The apparent pulse period is affected by a number of
phenomena, including the spin-down, timing noise, and/or glitches
intrinsic to the pulsar, variations in the interstellar dispersion,
and Doppler effects introduced by the relative motions of the Earth
and pulsar.  Inaccuracies in the model that describes these effects
introduce phase errors that accumulate with time and cause the
integrated profile to become smeared.  Therefore, it is often
beneficial to store multiple, shorter integrations of the mean pulse
profile instead of a single, long integration.  Furthermore, when a
pulsar is bright enough, a great deal of additional information about
the characteristics of the pulsar emission can be obtained by
recording and analysing each individual pulse.  Therefore, a useful
pulsar data format must be able to represent pulse profiles observed
over multiple epochs of arbitrary length.\\

In summary, pulsar observations generally consist of a
four-dimensional array of data indexed by polarization component,
pulse phase, frequency, and epoch.  Software support for sensible
groupings in other dimensions, such as orbital phase, is also highly
desirable.  In addition, data from a number of telescopes can be
combined to increase sensitivity and contribute to the eventual
detection of new phenomena, such as the cosmic background of
stochastic gravitational radiation (e.g. Hellings \& Downs 1983;
Stinebring et al. 1990).  Therefore, the data storage format should
have a flexible structure that provides efficient access to key
parameters, removed from any considerations of individual instruments
or signal processing schemes.

\subsection{Processing radio pulsar data}
\label{sec:processing}

Pulsars are observed for a variety of reasons, from studying the
nature of their structure and emission mechanism (Dodson, McCulloch,
\& Lewis 2002) to utilising them as highly stable clocks and
astrophysical probes (Taylor \& Weisberg 1982). Consequently, the same
pulsar observation can be used in a number of different contexts, one
focusing on the variation of polarization with frequency, another
measuring general relativistic effects on pulse times of arrival, etc.
Nevertheless, our experience has shown that there exist many common
tasks associated with pulsar data analysis that can be standardised
within a modern open source development environment.\\

As a demonstration of the types of operations performed on pulsar
data, consider the specific example of pulse time of arrival
calculation.  In order to increase the signal-to-noise ratio (S/N) of
each observation, data are often integrated (scrunched) by several
factors in one or more of the available dimensions.  Each resultant
profile is then cross-correlated with a high S/N standard profile
known as a template, yielding an estimate of the longitudinal offset
between the two.  This offset is added to the reference epoch
associated with a fiducial point in the observed pulse profile,
yielding an arrival time in the reference frame of the observatory,
which is later converted into a barycentric arrival time using a Solar
System ephemeris.  This data reduction operation involves a number of
typical tasks, including loading the arrays of numbers that represent
the folded profiles and computing sums, products, rotations, weighted
averages, and correlations of these arrays; sometimes in the Fourier
domain.  Most of these various operations must be performed in a
manner consistent with the observational parameters, taking into
account dispersive delays, observation time stamps and relative
weightings of different frequency channels, for example.  At each
step, the software must also ensure that all parameters are updated
accordingly.

\subsection{Scope and design of PSRCHIVE and PSRFITS}

It should be noted that the pulsar data under consideration represents
a point near the end of the typical pulsar data reduction chain. The
software presented in this paper is not intended for the direct
handling of radio data such as that recorded by baseband systems, or
for the purposes of performing computationally expensive offline
searching, although some support for the storage of such data is
provided in PSRFITS. The code is also not designed to perform any
phase-coherent dispersion removal or formation of filter-bank data;
these techniques are treated as separate computational tasks. Code for
such data reduction is also available from the repository at the
Swinburne Centre for Astrophysics and Supercomputing under the
umbrella name of BASEBAND
DSP\footnote{http://astronomy.swin.edu.au/pulsar/software/libraries/dsp},
a general library for use in digital signal processing.\\

The PSRCHIVE and PSRFITS schemes were designed from the beginning to
form an object-oriented framework into which existing algorithms and
data structures could be transplanted.  By introducing layers of
abstraction between the various levels of responsibility, the design
remains both flexible and extensible.  For example, different
telescopes and instruments require the storage of different types of
information, including configuration parameters, observatory and
instrumental status information, and other site-specific data.
Because there is no way of knowing exactly what future systems might
include, both PSRCHIVE and PSRFITS implement a generalised scheme for
incorporating arbitrarily complex data extensions, as described in
Sections~\ref{sec:abstraction} and~\ref{sec:psrfits_defn}.\\

In addition, a basic framework of crucial parameters common to all
pulsar observations and a wide variety of fundamental data reduction
algorithms, such as those described in Section~\ref{sec:processing},
have been implemented.  Each of these basic data structures and
reduction operations may be used in the composition of more complex
scientific processing algorithms.  By virtue of continued development
amongst the authors, the PSRCHIVE library includes an extensive array
of such high-level algorithms for use in the calibration,
visualisation, and analysis of pulsar data; these can be used
immediately on any of the supported file formats.\\

PSRCHIVE and PSRFITS were developed in parallel and are presented in
the hope that they will promote increased data portability.  The
PSRFITS file format also serves as an example of how to incorporate
other, pre-existing file formats into the new scheme, as described in
Section~\ref{sec:plugin}.  After two years of development, the code is
now ready for formal release to the wider pulsar community.  In the
following sections, we describe the implementation of the new schemes
and outline the specific advantages that they offer.

\section{Implementation Overview}

\subsection{Object-oriented programming}

The modularity and extensibility required of our new scheme suggested
an object-oriented approach. Since much of the existing Swinburne
analysis code had already been written in both the C and C++
programming languages, it seemed a natural step to progress in
C++. The concepts of object classes and inheritance provided and
enforced by the syntax of this language offer a sound foundation on
which to develop. In particular, object-oriented design has aided in
the realization of simultaneous support of multiple file formats. We
are aware that a majority of pulsar research groups prefer to write a
more procedural style of code, using FORTRAN or C.  However, we feel
that the benefits of an object-oriented approach to data processing
significantly outweigh the potential learning curve involved in
becoming proficient with C++.

\subsection{Basic class structure}

The required functionality of PSRCHIVE is built around a core
framework of C++ object classes. The fundamental unit of all pulsar
observations is the individual pulse \textbf{Profile}, a
one-dimensional array of floating point numbers, indexed by pulse
phase.  The \textbf{Integration} is a two dimensional vector of
\textbf{Profile} instances, indexed by frequency and polarisation, as
measured over a particular epoch.  In turn, the \textbf{Archive} is a
one dimensional vector of \textbf{Integration} instances, indexed in
one of a number of possible ways (normally time).  Each of these
classes implement a wide range of basic data manipulation and
processing operations.\\

In the language of C++, we define the namespace \textbf{Pulsar}, which
contains the three \textit{base classes}: \textbf{Pulsar::Archive},
\textbf{Pulsar::Integration}, and \textbf{Pulsar::Profile}.  In
addition, there are other object classes in the \textbf{Pulsar}
namespace that deal with specific tasks related to pulsar data
analysis. For example, the \textbf{Pulsar::Calibration} class employs
various mathematical models of the instrumental response to calibrate
polarimetric observations (van Straten 2004).

\subsection{Use of data abstraction}
\label{sec:abstraction}

The three base classes implement a wide variety of basic algorithms,
known as \textit{methods}, that are commonly used in pulsar data
analysis.  However, they do not require knowledge of any specific
details related to system architecture, enabling their use as
templates upon which to base lower-level development.  These templates
define the minimum set of parameters, known as \textit{attributes},
required to implement the data analysis methods, including
observational parameters such as the name of the source, centre
frequency, bandwidth, etc.  At the level of the
\textbf{Pulsar::Archive} and \textbf{Pulsar::Integration} base
classes, nothing is known about how data are stored on permanent media
or in computer memory.\\

\begin{figure}
\centerline{\psfig{width=15cm,figure=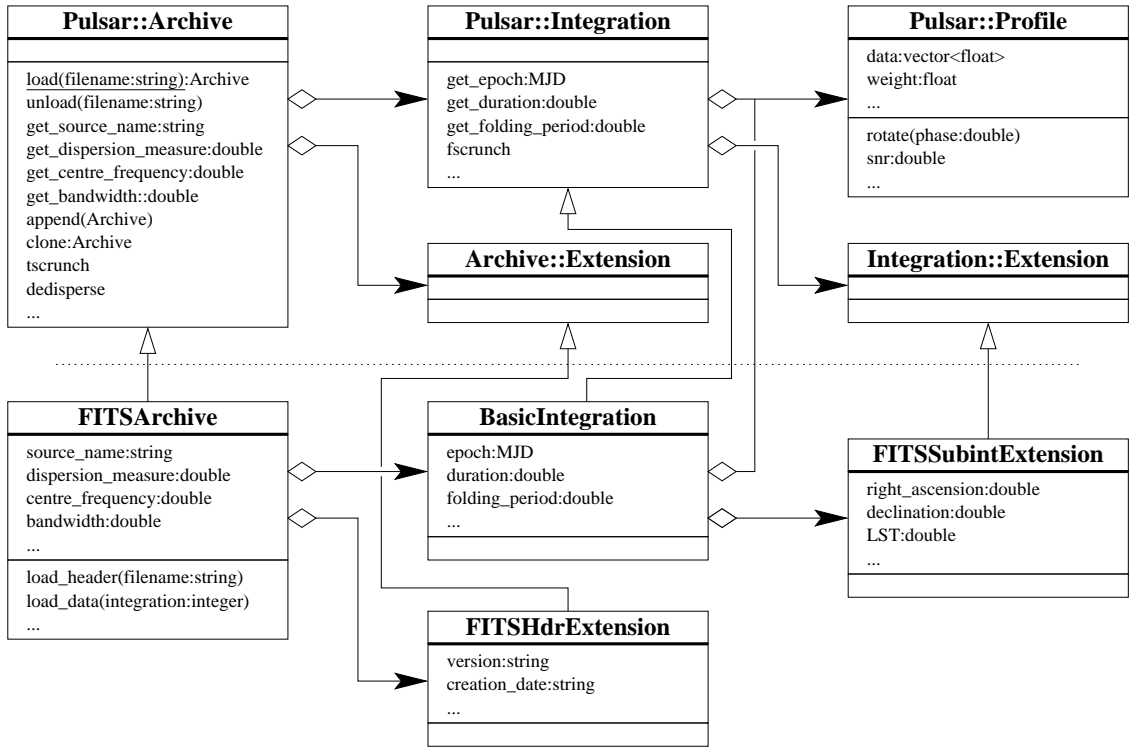}}
\caption {Class diagram of a portion of the PSRCHIVE library.  The
abstract base classes are shown above the dotted line. Below this
line, the \textbf{FITSArchive} class implements
\textbf{Pulsar::Archive} attribute storage and access methods, as well
as methods for loading and unloading data to and from a PSRFITS file.
The combined use of composition and inheritance enables complex
structures and behaviours to be constructed using modular
components. }
\label{fig:class_diagram}
\end{figure}

The necessary task of translating between the two realms is performed
by \textit{derived classes} that \textit{inherit} the base classes.
In order to inherit a base class, it is necessary for the derived
class to provide access to the required attributes and to implement
the methods used to read and write the data stored on disk.
Therefore, for each specific file format represented in the PSRCHIVE
scheme, there corresponds a derived class that inherits
\textbf{Pulsar::Archive}.  The syntax for the data access and file
input/output methods is defined by the base class and enforced by the
C++ compiler, allowing all derived classes to be treated as equal.
Therefore, high-level code can be written in the language of the base
class definition without the need for considering the implementation
details of the derived classes.  This abstraction, which is crucial to
the flexibility of the PSRCHIVE scheme, is demonstrated by the Unified
Modeling Language (UML) class diagram shown in
Figure~\ref{fig:class_diagram}.

\subsection{File format plug-in libraries}
\label{sec:plugin}

In order to take full advantage of this level of data abstraction, the
PSRCHIVE scheme makes use of dynamic shared object libraries, or
\textit{plug-ins}.  These libraries are compiled using special options
that allow them to be linked into a program at run-time.  Perhaps the
best known example of such a system is the plug-in scheme used to add
functionality to many web browsers.  Within PSRCHIVE, the machine code
that defines a \textbf{Pulsar::Archive}-derived class is stored inside
a single plug-in file.  The plug-in files corresponding to different
file formats are held in a central location that is scanned on the
first attempt to load pulsar data.  The available plug-ins determine
which file formats are understood by providing a test routine that
returns true if a file on disk is of its own format.  In this way,
PSRCHIVE applications can quickly scan a given data file, select the
appropriate derived class and load the pulsar data.  This ensures
that, as the number of supported file formats grows, the size and
complexity of any given application program remains the same. We 
encourage all interested research groups that currently maintain
separate data formats to gain experience with the PSRCHIVE scheme by
developing their own file format plug-ins.  By making the plug-in code
publicly available, research groups will be able to exchange data
already stored using different file formats.\\

In order to accommodate the stringent reliability requirements of
observatory online processing and monitoring systems, we also offer
the option of compiling all PSRCHIVE applications using a static
linkage scheme. This makes the binary executables larger on disk but
removes the possibility of problems arising should a shared object
file be accidentally deleted or recompiled. Either option can be
selected by simply editing one line in the Makefile scheme.

\section{PSRFITS}
\subsection {A standard format for pulsar data storage}

One of the motivating factors behind the development of the PSRCHIVE
scheme was the alleviation of problems associated with highly specific
and non-portable data storage formats.  This effort has highlighted
several compelling reasons for the pulsar community to move towards a
more modular and standardised storage format.  For instance, the
existence of a standard file format would significantly decrease the
amount of effort required to integrate and test new instrumentation.
Historically, file formats have accreted features as they became
desirable or necessary. Given the wealth of past experience available,
it seems a logical step to define a new format that encompasses a wide
range of features from the beginning and is written in a modular way
so as to enable rapid, backwards-compatible upgrades. Indeed, one
particular standard storage format has already won wide acclaim within
the astronomical community; the Flexible Image Transport System (FITS)
(Hanisch et al. 2001) has been in widespread use for approximately
three decades and has evolved into a highly adaptable data storage
scheme\footnote{http://archive.stsci.edu/fits/fits\_standard}. The
format has been placed under the administration of the IAU FITS
Working
Group\footnote{http://www.cv.nrao.edu/fits/traffic/iaufwg/iaufwg.html}
and a wide array of software is available for FITS file
manipulation. The NASA High Energy Astrophysics Science Archive
Research
Centre\footnote{http://heasarc.gsfc.nasa.gov/docs/heasarc/fits.html}
provides useful libraries and applications for manipulation and
interrogation of FITS-based files. For example, the program
\textbf{fv} has made the process of testing and debugging the relevant
PSRCHIVE software much more straight forward.\\

In accordance with FITS standards, a PSRFITS file consists of a
primary header-data unit (HDU) followed by a series of extension
HDUs. The primary HDU contains basic information such as telescope
identification and location, observation start time etc. Extension
HDUs, formatted as binary tables, contain specific information related
to the observation such as the pulsar ephemeris, calibration data, and
the pulsar data.  Although PSRFITS is primarily designed to store
folded or single-pulse profile data, it can also accommodate
continuous time series data.

A useful feature of the standard FITS input/output routines is that
new HDUs and header parameters may be added transparently -- if they
are unknown to the reading program, they are ignored. Furthermore,
unused HDUs need not be written, even though they are present in the
definition. This feature allows, for example, a user group to add
information particular to a certain instrument without compromising
use of the definition by other groups.

A novel feature of the PSRFITS definition is the inclusion of HDUs
containing `history' information. For example, the first line of
Processing History HDU contains information about the data acquisition
program and the initial structure of the file.  Subsequent lines
record details of modifications to the structure or data (for example,
partial or complete de-dispersion or interference excision).  A
permanent record of the steps that have been applied during data
reduction has proven to be of great value when later assessing the
quality and validity of observational data.

\subsection{The PSRFITS Definition}
\label{sec:psrfits_defn}

The current version of the PSRFITS definition file is available on the
ATNF web
pages\footnote{http://www.atnf.csiro.au/research/pulsar/psrfits}.
Table~\ref{tb:psrfits} describes the header-data units included in the
current definition (Version 1.18).\\

In addition to the Main and Processing History HDUs, a number of
optional HDUs have been defined for general use with a variety of
instrumentation. These enable the storage of important status and
diagnostic information about the observation, and demonstrate the
modularity and extensibility of the PSRFITS file format.  The physical
parameters stored in the Ephemeris History HDU are based on the pulsar
timing program,
\textbf{tempo}\footnote{http://www.atnf.csiro.au/research/pulsar/tempo}.
From the ephemeris parameters are derived the polynomial coefficients
(polyco) used to predict the apparent pulsar period and phase at the
epoch of the observation; these coefficients are stored in the Polyco
History HDU.  As improved physical parameters become available, the
data may be reprocessed, leading to new rows in the Ephemeris and
Polyco history tables.  The calibration and feed cross-coupling HDUs
are designed to work with the routines in the
\textbf{Pulsar::Calibration} class.  Owing to the intrinsic modularity
of FITS, these additional HDUs are all optional; in fact, it is not
even strictly necessary to include any Integration data in a PSRFITS
file.  For example, the polarimetric calibration modeling program
creates a file containing only the feed cross-coupling, injected
calibration polarisation, and flux calibration HDUs.  This modularity
is similar to that made available through the use of
VOTable\footnote{http://www.ivoa.net} XML standards and it is likely
that PSRFITS could in future be incorporated into the International
Virtual Observatory system with a minimum of effort.

\begin{table*}
\caption{PSRFITS -- A summary of the current definition}
\label{tb:psrfits}
\begin{tabular}{ll}\hline
HDU Title & \multicolumn{1}{c}{Description} \\ \hline
\\
Main header & Observer, telescope and receiver information, \\
            & source name and observation date and time \\
\\
Processing history & Date, program and details of data acquisition \\
            & and each subsequent processing step \\
\\
Digitiser statistics & Digitiser mode and count statistics \\
\\
Digitiser counts & Digitiser mode and count rate distribution \\
\\
Original bandpass & Observed bandpass in each polarisation \\
                  & averaged over the observation  \\
\\
Coherent de-dispersion & Parameters for coherent de-dispersion of \\
                      & baseband data \\
\\
Ephemeris history & Pulsar parameters used to create or modify \\
                  & profile data \\
\\
Polyco history & Elements of the polyco file used to predict \\
               & the apparent pulsar period \\
\\
Flux calibration & System temperature and injected noise \\
                 & calibration data as a function of frequency \\
                 & across the bandpass \\
\\
Injected calibration polarisation & Apparent polarisation of the injected \\
                 & noise calibration signal as a function \\
                 & of frequency\\
\\
Feed cross-coupling & Parameters of feed cross-coupling as a \\
                    & function of frequency \\
\\
Integration data & Pulse profiles or fast-sampled data as a function \\
                 & of time, frequency and polarisation \\
\\
\hline
\end{tabular}
\end{table*}


\section{Working With the PSRCHIVE Scheme}
\subsection{The standard application set}

The PSRCHIVE scheme includes an extensive set of pre-written
application programs that can be used to manipulate pulsar data in
various ways. These include both command line tools and graphical user
interfaces built using Trolltech's
Qt\footnote{http://www.trolltech.com/products/qt/index.html}, a C++
toolkit for multi-platform GUI and application development. Table
\ref{tb:psrapps} presents a list of applications included in the
package at the time of publication, with a brief description of each.

\begin{table*}
\caption{Standard applications included with PSRCHIVE}
\label{tb:psrapps}
\begin{tabular*}{1.0\textwidth}%
                {@{\extracolsep{\fill}}ll}\hline

Application & \multicolumn{1}{c}{Description} \\ \hline
\\
\textbf{pav}    & Archive data visualisation. Based on the PGPLOT graphics \\
                & subroutine library with a simple command line interface \\
\\
\textbf{vap}    & Archive header examination, allowing multiple user \\
                & selectable header parameters to be printed as ASCII to \\
                & the terminal \\
\\
\textbf{pam}    & Archive manipulation, compression and processing \\
\\
\textbf{pat}    & Pulse profile arrival time calculation, based on cross \\
                & correlation with a standard template profile \\
\\
\textbf{pas}    & Standard profile phase alignment, for timing with multiple \\
                & standard template profiles \\
\\
\textbf{paz}    & Radio frequency interference mitigation tool including \\
                & manual and automated channel zapping and sub-integration \\
                & removal \\
\\
\textbf{pac}    & Archive polarimetric and flux calibration tool based on a \\
                & user-selectable set of advanced algorithms \\
\\
\textbf{pcm}    & Polarimetric calibration modeling, creates instrumental \\
                & response transformations for use with \textbf{pac} \\
\\
\textbf{psrgui} & Interactive point-and-click data visualisation with a Qt \\
                & graphical interface \\
\\
\textbf{psradd} & Combination of multiple archives for formation of high \\
                & S/N profiles \\
\\
\textbf{rhythm} & A graphical interface for pulse arrival time fitting based \\
                & on \textbf{tempo} \\
\\
\hline
\end{tabular*}
\end{table*}

Readers may note that the modular philosophy at the heart of PSRCHIVE
extends all the way through to the user level applications. Each
program tends to be small and focused on a specific task, be it data
compression, timing, RFI mitigation, etc. This greatly simplifies
development and maintenance compared to having one monolithic program
with multiple purposes.

\subsection{PSRCHIVE as a development environment}

PSRCHIVE was designed to provide users with more than just a set of
pre-made applications. The classes, libraries and examples provided
are intended to simplify the task of building new processing tools. To
a large extent, developers who build on the PSRCHIVE scheme do not
have to directly manipulate the arrays of pulse profile
amplitudes. Instead, member functions of the various classes can be
called to perform basic operations like baseline removal and phase
rotation. This has the dual benefit of labour saving both in the
initial development phase and in the debugging phase, as both the
authors and other users have already verified and tested the provided
routines. In case direct access to the profile amplitudes is required,
we also provide interface functions that return C style arrays. In the
experience of the authors, the extra layer of abstraction provided by
the PSRCHIVE scheme can cut down the time between program concept and
full implementation to a matter of hours. New applications can be
built with only a few lines of code. For example, to remove the system
noise floor, compress all frequency channels and output the processed
archive:

\begin{verbatim}
# include "Pulsar/Archive.h"

int main() {

  Pulsar::Archive* arch = 0;

  arch = Pulsar::Archive::load("filename");

  arch->remove_baseline();
  arch->fscrunch();

  arch->unload();

}
\end{verbatim}

This simple program defines a pointer to a \textbf{Pulsar::Archive}
and calls the generic \textbf{Pulsar::Archive::load} routine, which
takes a filename argument, applies a number of tests to the file on
disk (depending on the available plug-ins) and decides whether or not
it understands the particular format. If so, it summons the
appropriate derived class to read the data from disk. Once the data
have been loaded, the \textbf{Pulsar::Archive::remove\_baseline}
function is called.

\begin{verbatim}
void Pulsar::Archive::remove_baseline (float phase, float width)
{
  try {

    if (phase == -1.0)
      phase = find_min_phase (width);

    for (unsigned isub=0; isub < get_nsubint(); isub++)
      get_Integration(isub) -> remove_baseline (phase, width);

  }
  catch (Error& error) {
    throw error += "Pulsar::Archive::remove_baseline";
  }
}
\end{verbatim}

The \textbf{Pulsar::Archive::remove\_baseline} function takes two
arguments: the {\tt phase} and {\tt width} of the off-pulse baseline.
Both arguments are assigned default values in the {\tt Archive.h}
header file; if {\tt phase} is left unspecified, then the off-pulse
baseline phase will be found using the
\textbf{Pulsar::Archive::find\_min\_phase} method; if {\tt width} is
unspecified, then a default value will be used.  The
\textbf{Pulsar::Archive::remove\_baseline} method makes multiple calls
to the \textbf{Pulsar::Integration::remove\_baseline} routine, which
performs the actual modification of amplitudes as follows:

\begin{verbatim}
void Pulsar::Integration::remove_baseline (float phase, float width)
{

  if (Pulsar::Integration::verbose)
    cerr << "Pulsar::Integration::remove_baseline entered" << endl;

  try {

    if (phase == -1.0)
      phase = find_min_phase (width);

    vector<float> phases;
    dispersive_phases (this, phases);

    for (unsigned ichan=0; ichan<get_nchan(); ichan++) {

      float chanphase = phase + phases[ichan];

      for (unsigned ipol=0; ipol<get_npol(); ipol++)
        *(profiles[ipol][ichan]) -=
          profiles[ipol][ichan] -> mean (chanphase, width);

    }

  }
  catch (Error& error) {
    throw error += "Integration::remove_baseline";
  }
}
\end{verbatim}

This nested structure reduces the length of high-level routines,
allowing actual computations to be done at the level of abstraction
that best suits the task.  Likewise, the
\textbf{Pulsar::Integration::remove\_baseline} routine calls various
member functions of both the \textbf{Pulsar::Integration} and
\textbf{Pulsar::Profile} classes, computing the pulse phase at which
the minimum baseline level occurs in the total intensity of the entire
band. Adjustments for dispersive delays in each channel are performed
and the mean level at this phase is individually removed from each
\textbf{Pulsar::Profile} stored in the
\textbf{Pulsar::Integration}. Developers should also note the
extensive use of \textbf{try/catch} blocks and a specifically designed
\textbf{Error} class that carries descriptive information about any
exceptions thrown back to the calling procedure.

\section{Resources and Availability}
\subsection{Obtaining and compiling the code}

PSRCHIVE is freely available to the worldwide academic community. It
is held in a repository at Swinburne University of Technology in
Melbourne, Australia and may be accessed via the Concurrent Versions
System\footnote{http://www.cvshome.org/}.  As it is distributed as
source code, some experience with programming and compilation is
necessary; however, installation can be done in a fairly simple
step-by-step manner thanks to the standard Makefile scheme included
with the package. The code is compatible with all versions of the GNU
Compiler Collection\footnote{http://gcc.gnu.org/} between 2.95 and
3.2.2 and is routinely tested on both the Linux and Solaris operating
systems.  Every effort will be made to ensure compatibility with
future gcc releases.\\

The PSRCHIVE scheme makes use of several external libraries, including
the Starlink Project\footnote{http://www.starlink.rl.ac.uk/} SLALIB package.
It also requires at least
one external FFT library and includes wrappers that provide
compatibility with either FFTW 2.1.5\footnote{http://www.fftw.org}
(available under the GPL) or Intel
MKL\footnote{http://www.intel.com/software/products/mkl/}
(commercially available from Intel). The
PGPLOT\footnote{http://www.astro.caltech.edu/\~tjp/pgplot/} graphics
subroutine library is also an integral part of the scheme.\\

Full documentation including instructions for download and
installation are available online by following the menu options at the
Swinburne Centre for Astrophysics and Supercomputing web
site\footnote{http://astronomy.swin.edu.au/pulsar/}.  Read-only access
to the repository is granted upon receipt by the developers of a
Secure Shell v2.0 public key that is used to allow remote entry to the
server. Write permissions to the repository require a computing
account with the Swinburne Centre for Astrophysics and
Supercomputing. Please direct all enquiries regarding access to
\textbf{psrchive@astro.swin.edu.au}.

\subsection{Online documentation}

PSRCHIVE reference documentation is maintained online.  In addition to
the online descriptions, each command line application has a
\textbf{-h} option that displays a quick summary of how to use the
program.  The library of C++ classes is extensively documented using
the Doxygen\footnote{http://www.doxygen.org} system; the source code
contains tagged comments from which the online manual is automatically
generated. This manual is intended as a reference to programmers as it
primarily describes the member functions available in each class and
the syntax of their arguments.

\subsection{Support services}

Although we provide no official support for the software, we are
willing to assist with PSRCHIVE related problems as time
permits. General queries regarding installation or operation can be
addressed to \textbf{psrchive@astro.swin.edu.au}.  We also provide a
mechanism for reporting serious bugs via an online interface known as
YAQ, which can be found at the Swinburne pulsar group web site.

\section{Conclusion}

The task of organising astronomical data into a logical format lends
itself surprisingly well to the object-oriented programming paradigm.
The combination of PSRCHIVE and PSRFITS provides a powerful,
ready-to-use pulsar data archival and reduction system that can be
rapidly adapted to new instruments.  We hope that the ready
availability of an open source data reduction framework will
facilitate large scale collaborative projects, such as an extended
pulsar timing array (Foster and Backer 1990).  Therefore, we encourage
both scientists and engineers involved with pulsar data acquisition
and reduction to consider taking advantage of these packages.

\section*{Acknowledgments}

We have benefited greatly from the advice and assistance of many
colleagues in developing the scheme described. In particular, we thank
Matthew Bailes and others in the Swinburne pulsar group and Nina Wang
at the Australia Telescope National Facility. Thanks also to Ben
Stappers, Russell Edwards and George Hobbs for constructive feedback
during development. We would also like to thank the referee for
insightful comments which have led to important improvements to this
manuscript.


\section*{References}






\reference Dodson, R. G., McCulloch, P. M. \& Lewis, D. R. 2002, ApJ, 564, 85

\reference Foster, R. S. \& Backer, D. C. 1990, ApJ, 361, 300

\reference Gold, T. 1968, Nature, 218, 731

\reference Hanisch, R. J., Farris, A., Greisen, E. W., Pence, W. D.,
Schlesinger, B. M., Teuben, P. J., Thompson, R. W. \& Warnock, A., III
2001, A\&A, 376, 359 

\reference Helfand, D. J., Manchester, R. N. \& Taylor, J. H. 1975,
ApJ, 198, 661

\reference Hellings, R. W. \& Downs, G. S. 1983, ApJL, 265, 39

\reference van Straten, W. 2004, ApJS, in press

\reference Stinebring, D. R., Ryba, M. F., Taylor, J. H. \& Romani,
R. W. 1990, Phys. Rev. Lett., 65, 285

\reference Taylor, J. H. \& Cordes, J. M. 1993, ApJ, 411, 674

\reference Taylor, J. H. \& Weisberg, J. M. 1982, ApJ, 253, 908


\end{document}